\documentstyle{article}
\begin{document}
\begin{center}
{\bf FURTHER SOLUTIONS OF FRACTIONAL REACTION-DIFFUSION EQUATIONS
IN TERMS OF THE H-FUNCTION}\\[0.5cm]
H.J. HAUBOLD\\
Office for Outer Space Affairs, United Nations\\
P.O. Box 500, A–1400, Vienna, Austria\\
and Centre for Mathematical Sciences, Pala-686 574, Kerala State, India\\[0.3cm]
A.M. MATHAI\\
Department of Mathematics and Statistics, McGill University\\
Montreal, Canada H3A 2K6\\
and Centre for Mathematical Sciences, Pala-686 574, Kerala State, India\\[0.3cm]
R.K. SAXENA\\
Department of Mathematics and Statistics, Jai Narain Vyas University\\
Jodhpur-342004, India
\end{center}
\noindent
Abstract. This paper is a continuation of our earlier paper in which we have derived the solution of an unified fractional reaction-diffusion equation associated with the Caputo derivative as the time-derivative and the Riesz-Feller fractional derivative as the space-derivative. In this paper, we consider an unified reaction-diffusion equation with Riemann-Liouville fractional derivative as the time-derivative and Riesz-Feller derivative as the space–derivative. The solution is derived by the application of the Laplace and Fourier transforms in a compact and closed form in terms of the H-function. The results derived are of general character and include the results investigated earlier by Kilbas et al. (2006a), Saxena et al. (2006c), and Mathai et al. (2010). The main result is given in the form of a theorem.  A number of interesting special cases of the theorem are also given as corollaries.

\section{Introduction}
In recent years, fractional reaction-diffusion models are studied due to their usefulness and importance in many areas of mathematics, statistics, physics, and chemistry (Mainardi, 2010; Mathai and Haubold, 2008; Haubold and Mathai, 2010; Mathai et al. 2010). Such models, formulated in standard or fractional calculus, greatly contribute to the understanding of the behavior of many-body systems far from equilibrium and emerging spatio-temporal pattern formation. Recently, coupled fractional reaction–diffusion equations are solved by Gafiychuk et al. (2006). Turing pattern formation through linear stability analysis and numerical simulation are discussed by Langlands et al. (2007). Nonlinear oscillations and stability domains in fractional diffusion systems with two types of variables activator and inhibitor are demonstrated by Gafiychuk et al. (2007). General models for reaction-diffusion systems are investigated by Henry and Wearne(2000, 2002) and Henry et al. (2005). 

This paper deals with the investigation of the solution of an unified model of reaction–diffusion system associated with the Riemann-Liouville fractional derivative as the time-derivative and the Riesz-Feller derivative as the space-derivative. This new model provides the extension of the models discussed earlier by Kilbas et al. (2006a), Saxena et al. (2006c), and Mathai et al. (2010). As special cases of this general model, we discuss neutral fractional diffusion model, time-fractional diffusion model and space-time fractional diffusion model. The present study is in continuation of our earlier works, Haubold and Mathai (1995, 2000, 2008, 2010), Haubold et al. (2007, 2010) and  Saxena et al. (2006a, 2006b, 2006c).

\section{Results Required in the Sequel}
 
 The Riemann–Liouville fractional integral of order $\nu$ is defined by Miller and Ross (1993, p. 45) and  Kilbas et al. (2006) 
\begin{equation}
_0D_t^{-\nu}N(x,t)=\frac{1}{\Gamma(\nu)}\int^t_0(t-u)^{\nu-1}N(x,u)du,
\end{equation}
where $Re(\nu)>0.$

 The Riemann-Liouville fractional derivative of order $\alpha>0$ is defined as (Samko et al., 1990, p.37; see also Kilbas et al., 2006) 
\begin{equation}
_0D_t^\alpha f(x,t)=\frac{1}{\Gamma(n-\alpha)}\frac{d^n}{dt^n}\int^t_0\frac{f(x,\tau)d\tau}{(t-\tau)^{\alpha+1-n}}\;\;(n=[\alpha]+1),\;n\in N.\;t>0
\end{equation}
where $[\alpha]$ means the integral part of the number $\alpha$.

	The Laplace transform of the  Riemann-Liouville fractional derivative  is given by Oldham and Spanier (1974, eq.(3.1.3); see also Kilbas et al., 2006)
\begin{equation}
L\left\{_0D^\alpha_t N(x,t); s\right\}=s^\alpha N(x,s)-\sum^n_{r=1}s^{r-1}\;_0D_t^{\alpha-r}\;N(x,t)|t=0 (n-1<\alpha\leq n).
\end{equation}
The Riesz-Feller space-fractional derivative of order $\alpha$ and skewness $\theta$ is defined in terms of its Fourier transform as (Feller, 1952, 1971):
\begin{equation}
F\left\{_xD_\theta^\alpha f(x); k\right\}= - \Phi^\theta_\alpha(k)f^*(k),
\end{equation}
where $f^*(k)$ denotes the Fourier transform of $f(t)$, defined by
\begin{equation}
f^*(k)=\int^\infty_{-\infty}exp(-ikt)f(t)dt
\end{equation}
and
\begin{equation}
\Phi_\alpha^\theta (k)=|k|^\alpha exp[i(sign k)\frac{\theta \pi}{2}], \;0<\alpha\leq 2,\;\;|\theta|\leq\; min\left\{\alpha,2-\alpha\right\}.
\end{equation}           		
When $\theta=0$, we have a symmetric operator with respect to $x$ that can be interpreted as
\begin{equation} 
_xD_0^\alpha = -\left(-\frac{d^2}{dx^2}\right)^{\alpha/2}.
\end{equation}     
This can be formally deduced by writing $-(k)\alpha=-(k^2)^{\alpha/2}$. Eq. (4) then reduces to 
\begin{equation}
F\left\{_xD_0^\alpha f(x); k\right\}= - |k|^\alpha,
\end{equation}
which is the Fourier transform of the Weyl fractional operator, defined by
\begin{equation}
_{-\infty} D_x^\mu f(t)=\frac{1}{\Gamma(n-\mu)}\frac{d^n}{dt^n}\int^t_{-\infty}\frac{f(u)du}{(t-u)^{\mu-n+1}}.
\end{equation}
This shows that Riesz-Feller operator may be regarded as a generalization of Weyl operator. An alternative notation for the symmetric fractional derivative $_\theta D_0^\alpha$ has been given by Saichev and Zaslavsky (1997) in the form                          
\begin{equation}
_\theta D_0^\alpha = \frac{d^\alpha}{d|x|^\alpha}.
\end{equation}            
In its regularized form, which holds for $0<\alpha<2$, the Riesz–Feller derivative admits the explicit representation
\begin{equation}
_xD_0^\alpha f(x)=\frac{\Gamma(1+\alpha)}{\pi} sin\left(\frac{\alpha \pi}{2}\right)\int_0^\infty\frac{f(x+\xi)-2f(x)+f(x-\xi)}{\xi^{1+\alpha}}d\xi.
\end{equation}
For $\alpha=1$, the Riesz derivative is related to the Hilbert transform as pointed out by Feller (1952). We have 
\begin{equation}
_xD_\eta^1 f(x)=-\frac{1}{\pi}\frac{d}{dx}\int^\infty_{-\infty} \frac{f(\xi)d\xi}{x-\xi}.
\end{equation}
For $0<\alpha<2$ and $|\eta|\leq min\left\{\alpha, 2-\alpha\right\}$, the Riesz-Feller derivative can be shown to possess the following integral representation in $x$ domain:
\begin{eqnarray}
&&_xD_\theta^\alpha f(x)\nonumber\\
&&=\frac{\gamma(1+\alpha)}{\pi}\left\{sin[(\alpha+\theta)\pi/2] \int^\infty_0\frac{f(x+\xi)-f(x)}{\xi^{1+\alpha}}d\xi+sin[(\alpha-\theta)\pi/2]\right.\nonumber\\
&&\left.\int^\infty_0\frac{f(x-\xi)-f(x)}{\xi^{1+\alpha}}d\xi\right\}
\end{eqnarray}                                                                                		                                
\section{Unified Fractional Reaction-Diffusion Equation}
In this section, we will investigate the solution of the reaction-diffusion system (14) under the initial conditions (15).\par
\smallskip
\noindent 
{\bf Theorem.}  Consider the unified fractional reaction-diffusion model associated with Riemann-Liouville fractional  derivative $_0D_t^\alpha$   defined by  (2) and the Riesz-Feller space fractional derivative $_xD_\theta^\alpha$   of order $\alpha$   and asymmetry $\theta$   defined by (4)
\begin{equation}
_0D_t^\beta N(x,t)=\eta_xD_\theta^\alpha N(x,t)+\Phi(x,t),
\end{equation}
where $\eta,t>0, x\in R; \alpha, \theta, \beta$  are real parameters with the constraints
$$0<\alpha\leq 2,|\theta|\leq min(\alpha, 2-\alpha), 1<\beta\leq 2,$$
and the initial conditions
\begin{equation}
_0D_t^{\beta-1}N(x,0)=f(x), \;_0D_t^{\beta-2} N(x,0)=g(x)\;\;\mbox{for}\; x\in R, \lim_{x \to \pm \infty}N(x,t)= 0, t>0.
\end{equation}
Here $_0D_t^{\beta-1} N(x,0)$   means the Riemann-Liouville fractional partial  derivative of  $N(x,t)$ with respect to $t$ of order $\beta-1$ evaluated at $t=0$. Similarly $_0D_t^{\beta-2}N(x,0)$ is the Riemann-Liouville fractional partial derivative of  $N(x,t)$  with respect to $t$ of order $\beta-2$   evaluated  at $t=0$. In (13), $\eta$   is a diffusion constant and $\Phi(x,t)$  is a nonlinear function belonging to the area of reaction-diffusion. Then for the solution of (14), subject to the above constraints, there holds the formula 
\begin{eqnarray}
N(x,t)&=&\frac{t^{\beta-1}}{2\pi}\int^\infty_{-\infty}f^*(k)E_{\beta, \beta}(-\eta t^\beta \psi^\theta_\alpha (k))exp(-ikx)dk\\
&+& \frac{t^{\beta-2}}{2\pi}\int^\infty_{-\infty}tg^*(k)E_{\beta,\beta-1}(-\eta t^\beta\psi^\theta_\alpha(k))exp(ikx)dk\nonumber\\
&+& \frac{1}{2\pi}\int^t_0\varsigma^{\beta-1}d\xi\int^\infty_{-\infty}\varphi^*(k,t-\xi)E_{\beta,\beta}(-\eta\xi^\beta\psi^\theta_\alpha(k))exp(-ikx)dk,\nonumber
\end{eqnarray}
where $E_{\alpha,\beta}(z)$ is the generalized Mittag-Leffler, defined by the series
$$E_{\alpha,\beta}(z)=\sum^\infty_{n=0}\frac{z^n}{\Gamma(\alpha n+\beta)}\;(\alpha,\beta) \in C, Re(\alpha)>0, Re(\beta)>0.$$	  
{\bf Proof.} If we apply the Laplace transform with respect to the time variable t, Fourier transform with respect to space variable $x$ and use the initial conditions (15) and the formula (17), then the given equation transforms into the form
$$s^\beta N^{^\sim_*}(k,s)-f^*(k)-sg^*(k)=-\eta\psi^\theta_\alpha(k)N^{^\sim_*}(k,s)+\varphi^{*\sim_*}(k,s),$$
where according to the conventions followed, the symbol $\sim$ will stand for the Laplace transform with respect to time variable $t$ and  * represents the Fourier transform with respect to space variable $x$. 
Solving for $N^{^*_\sim}(k,s)$, it yields
\begin{equation}
N^{^*_\sim}(k,s)= \frac{f^*(k)}{s^\beta+\eta\psi^\theta_\alpha(k)}+\frac{sg^*(k)s^{\beta-2}}{s^\beta+\eta\psi^\theta_\alpha(k)}+\frac{\phi^{^*_\sim}(k)}{s^\beta+\eta\psi^\theta_\alpha(k)}.
\end{equation}
On taking the inverse Laplace transform of (16) by using the formula 
\begin{equation}
L^{-1}\left\{\frac{s^{\beta-1}}{a+s^\alpha}\right\}=t^{\alpha-\beta}E_{\alpha,\alpha=\beta+1}(-at^\alpha),
\end{equation}
where   $Re(s) > 0, Re(\alpha)>0, Re(\alpha-\beta)>-1$; it is seen that 
\begin{eqnarray}
N^*(k,t)&=&t^{\beta-1}f^*(k)E_{\beta,\beta}(-\eta t^\beta\psi^\theta_\alpha(k))+t^{\beta-2}g^*(k)t E_{\beta,\beta-1}(-\eta t^\beta \psi^\theta\alpha(k))\nonumber\\
&+&\int_0^t\varphi^*(k,t-\xi)\xi^{\beta-1}E_{\beta,\beta}(-\eta\psi^\theta_\alpha(k)\xi^\beta)d\xi.
\end{eqnarray}
The required solution (16) is now obtained by taking the inverse Fourier transform of (19). This completes the proof of the theorem.

\section{Special Cases}
If we set $\theta = 0$ by virtue of the results (6), the theorem reduces to the following

{\bf Corollary 1.}   Consider the unified fractional reaction-diffusion model associated with Riemann-Liouville fractional  derivative $_0D_t^\alpha$  defined by  (2) and the Riesz-Feller space fractional derivative $_xD_0^\alpha$of order $\alpha$   defined by (11)
\begin{equation}
_0D_t^\beta N(x,t)=\eta_xD_0^\alpha N(x,t)+\varphi(x,t),
\end{equation}                                        		
where $\eta,t>0, x\in R; \alpha,\beta$  are real parameters with the constraints
$$0<\alpha<2, 1<\beta\leq 2,$$
and the initial conditions
\begin{equation} 
_0D_t^{\beta-1}N(x,0)= f(x), \;_0D_t^{\beta-2}N(x,0)=g(x)\;\;\mbox {for} x \in R, \lim_{x \to \pm \infty}N(x,t)=0, t>0.
\end{equation}
Here $_0D_t^{\beta-1} N(x,0)$   means the Riemann-Liouville fractional partial  derivative of  $N(x,t)$ with respect to $t$ of order $\beta-1$ evaluated at $t=0$. Similarly $_0D_t^{\beta-2}N(x,0)$ is the Riemann-Liouville fractional partial derivative of  $N(x,t)$  with respect to $t$ of order $\beta-2$   evaluated  at $t=0$. The $\eta$  is a diffusion constant and $\varphi(x,t)$ is a nonlinear function belonging to the area of reaction-diffusion. Then for the solution of (14), subject to the above constraints, there holds the formula 
\begin{eqnarray}
N(x,t)&=&\frac{t^{\beta-1}}{2\pi}\int_{-\infty}^\infty f^*(k)E_{\beta, \beta}(-\eta t^\beta|k|^\alpha)exp(-ikx)dk\nonumber\\
&+&\frac{t^{\beta-2}}{2\pi}\int^\infty_{-\infty}g^*(k)E_{\beta, \beta-1}(-\eta t^\beta|k|^\alpha)exp(-ikx)dk\nonumber\\
&+&\frac{1}{2\pi}\int^t_0\varsigma^{\beta-1} d\xi\int^\infty_{-\infty}\varphi^*(k,t-\xi)E_{\beta,\beta}(-\eta\xi^\beta|k|^\alpha)exp(-ikx)dk,
\end{eqnarray}
When $g(x) = 0$, then by the application of the convolution theorem of the Fourier transform to the solution (16) of the theorem, it readily gives\par
\medskip
\noindent  
{\bf Corollary 2.} The solution of fractional reaction–diffusion equation 
\begin{equation}
_0D_t^\beta N(x,t)-\eta\;_xD_\theta^\alpha N(x,t)=\varphi(x,t), x\in R, \;t>0, \eta>0,
\end{equation}
with initial conditions  
\begin{eqnarray}
&&_0D_t^{\beta-1}N(x,0)=f(x),\;_0D_t^{\beta-2}N(x,0)=0,\nonumber\\
&&\mbox{for}\; x\in R, -\leq\alpha\leq1, 1<\beta\leq 2, \lim_{x \to \pm\infty}N(x,t)=0,
\end{eqnarray}
where $\eta$ is  a diffusion constant and $\varphi(x,t)$  is a nonlinear function belonging to the area of reaction-diffusion; $\eta,t>0, x\in R; \alpha, \theta,\beta$  are real parameters with the constraints
$$0<\alpha\leq 2, |\theta|\leq min(\alpha, 2-\alpha), 1<\beta\leq 2$$
is given by 
\begin{eqnarray}
N(x,t)  &=&  \int_0^\alpha G_1(x-\tau,t)f(\tau)d\tau\nonumber\\
&+& \int_0^td\xi(t-\xi)^{\beta-1}\;\;\int_0^\alpha G_2(x-\tau, t-\xi)\varphi(\tau, \xi)d\tau,
\end{eqnarray}  
where
\begin{eqnarray}
\rho&=&\frac{\alpha-\theta}{2\alpha}, \nonumber\\
G_1(x,t)&=&\frac{t^{\beta-1}}{2\pi}\int^\infty_{-\infty} exp(-ikx)E_{\beta, \beta}(-\eta t^\beta\psi^\theta_\alpha(k))dk\nonumber\\
&=&\frac{t^{\beta-1}}{\alpha|x|}H^{2,1}_{3,3}\left[\frac{|x|}{\eta^{1/\alpha}t^{\beta/\alpha}}\left|^{(1,1/\alpha), (\beta, \beta/\alpha),(1,\rho)}_{(1,1),(1,1),(1,\rho)}\right.\right],\;(\alpha>0)
\end{eqnarray}  
and  
\begin{eqnarray}
G_2(x,t)&=&\frac{1}{2\pi}\int^\infty_{-\infty}exp(-ikx)E_{\beta,\beta}(-\eta t^\beta \psi^\theta_\alpha(k))dk\nonumber\\
&=&\frac{1}{\alpha|x|}H^{2,1}_{3,3}\left[\frac{|x|}{\eta^{1/\alpha}t^{\beta/\alpha}}\left|^{(1,1/\alpha), (\beta, \beta/\alpha), (1,\rho)}_{(1,1/\alpha), (1,1),(1,\rho)}\right.\right](\alpha>0).
\end{eqnarray}
Here $H^{2,1}_{3,3}(.)$  is the H-function , defined in Mathai and Saxena (1978, p.2; see also Mathai et al. 2010). 
In deriving the above results, we have used the inverse Fourier transform formula (Haubold et al., 2007)
\begin{equation}
F^{-1}\left[E_{\beta,\gamma}(-\eta t^\beta \psi^\alpha _\theta(k)); x)\right]=\frac{1}{\alpha|x|} H^{2,1}_{3,3}\left[\frac{x}{\eta^{1/\alpha}t^{\beta/\alpha}}\left|^{(1,1/\alpha), (\gamma,\beta/\alpha), (1,\rho)}_{(1,1/\alpha), (1,1),(1,\rho)}\right.\right],
\end{equation}        
where $Re(\alpha)>0 \;Re(\beta)>0,\;Re(\gamma)>0.$\\

It is interesting to observe that for $\theta=0$ Corollary 1 reduces to a result given by the authors (Saxena et al., 2006c).
On the other hand if we set $f(x)=\sigma(x),$   where $\sigma (x)$   is the Dirac-delta function, it yields\par
\medskip
\noindent
{\bf Corollary 3.} Consider the following reaction-diffusion model 
\begin{equation}
_0D_t^\beta N(x,t)=\eta\;_xD_\theta^\alpha N(x,t),
\end{equation}
with the initial conditions $_0D_t^{\beta-1}N(x,0)=\sigma(x), 0\leq\beta\leq 1, \lim_{x\to \pm\infty} N(x,t)=0,$ where $\eta$ is a diffusion constant; $\eta,t>0,x\in R; \alpha, \theta, \beta$  are real parameters with the constraints
$$0<\alpha \leq 2, |\theta| \leq min(\alpha, 2-\alpha),$$         
and $\delta(x)$ is the Dirac-delta function. Then for the fundamental solution of (29) with initial conditions, there holds the formula 
\begin{equation}
N(x,t)=\frac{t^{\beta-1}}{\alpha|x|}H^{2,1}_{3,3}\left[\frac{|x|}{(\eta t^\beta)^{1/\alpha}}\left|^{(1,1/\alpha), (\beta,\beta/\alpha), (1, \rho)}_{(1,1/\alpha), (1,1), (1,\rho)}\right.\right], (\alpha>0)
\end{equation}
where $\rho=\frac{\alpha-\theta}{2\alpha}.$
For $\theta=0$ we obtain the result given by Kilbas et al. (2006a) obtained in a different form. In this case the authors have given the result in a closed form (Saxena et al., 2006c, p.309).\par
\medskip
\noindent
{\bf Remark.} We note that the equation (29), when the Riemann-Liouville operator appearing on its left is replaced by a Caputo derivative (Caputo, 1969; Mainardi, 2010), has been solved by Mainardi et al. (2001, 2005).
  
The following special cases of (29) are worth mentioning:

(i)  For $\alpha=\beta$,  the corresponding solution of (29), denoted by $N_\alpha^\theta$,  we call as the neutral fractional diffusion, which can be expressed in terms of the H-function as given below and can be defined for $x>0$:\par
\medskip
\noindent
Neutral fractional diffusion: $0<\alpha=\beta<2; \theta\leq min \left\{\alpha,2-\alpha\right\},$
\begin{equation}
N_\alpha^\theta(x)=\frac{t^{\alpha-1}}{\alpha|x|}H^{2,1}_{3,3}\left[\frac{|x|}{t\eta^{1/alpha}}\left|^{
(1,1/\alpha),(\alpha,1), (1,\rho)}_{(1,1/\alpha),(1,1), (1,\rho)}\right]\right.,\; \rho=\frac{\alpha-\theta}{2\alpha}.
\end{equation}		
Next we derive some stable densities in terms of the H-function as special cases of the solution of the equation (26).
(ii) When $\beta=1,0<\alpha\leq2;\theta\leq min\left\{\alpha, 2-\alpha\right\}$then (29) reduces to space-fractional diffusion equation, which we denote by is the fundamental solution of the following space-time fractional diffusion model:
\begin{equation}
\frac{\partial N(x,t)}{\partial t}= \eta\;_xD_\theta^\alpha N(x,t), \eta>0, x\in R
\end{equation}
with the initial conditions  $N (x,t=0) = \sigma(x), \lim_{x\to\pm\infty} N(x,t)=0,$ where $\eta$ is a diffusion constant  and $\delta(x)$ is the Dirac-delta function. Hence for the solution of (29) there holds the formula 
\begin{equation}
L_\alpha^\theta(x)=\frac{1}{\alpha(\eta t)^{1/\alpha}}\;H^{1,1}_{2,2}\left[\frac{(\eta t)^{1/\alpha}}{|x|}\left|^{(1,1),(\rho, \rho)}_{(\frac{1}{\alpha},\frac{1}{\alpha}),(\rho, \rho)}\right]\right.,\;0<\alpha<1, |\theta|\leq \alpha,
\end{equation}       
where $\rho=\frac{\alpha-\theta}{2\alpha}$. The density represented by the above expression is known as $\alpha$-stable L\'{e}vy density. Another form of this density is given by 
     \begin{equation}
L_\alpha^\theta (x)=\frac{1}{\alpha(\eta t)^{1/\alpha}}\;H^{1,1}_{2,2}\left[\frac{|x|}{(\eta t)^{1/\alpha}}\left|^{(1-\frac{1}{\alpha},\frac{1}{\alpha}), (1-\rho, \rho)}_{(0,1),(1-\rho, \rho)}\right.\right], 1<\alpha 2, |\theta|\leq 2-\alpha.
\end{equation} 
{\bf Note:} A comprehensive account of stable densities with applications is available from the monograph of Uchaikin and Zolotarev (1999).

(iii) Next, if we take $\alpha=2,0<\beta<2;\theta =0$,  then we obtain the time-fractional diffusion,  which is governed by the following time-fractional diffusion model:
\begin{equation}
\frac{\partial ^\beta N(x,t)}{\partial t^\beta} = \eta \frac{\partial^2}{\partial x^2} N(x,t), \eta>0, x\in R,\; 0<\beta\leq 2,
\end{equation}
       with the initial conditions $_0D_t^{\beta-1} N(x,0)= \sigma(x), _0D_t^{\beta-2} N(x,0)=0,\; \mbox{for}\; x \in r, \lim_{x\to\pm\infty} N(x,t)=0$
where $eta$  is a diffusion constant  and $\sigma(x)$ is the Dirac-delta function, whose fundamental solution is given by the equation 
\begin{equation}
N(x,t)=\frac{t^{\beta-1}}{2|x|}\;H^{1,0}_{1,1}\left[\frac{|x|}{(\eta t^\beta)^{1/2}}\left|^{(\beta, \beta/2}{(1,1)}\right.\right].
\end{equation}
Further, if we set $\alpha=2, \beta=1$ and $\theta\rightarrow 0$, then for the fundamental solution of the standard diffusion equation
\begin{equation}
\frac{\partial}{\partial t}N(x,t)=\eta\frac{\partial^2}{\partial x^2}N(x,t),       
\end{equation}
with initial condition 
\begin{equation}
        N(x,t=0) = \sigma(x), \lim_{x\to \pm\infty} N(x,t)=0,
\end{equation}
there holds  the formula 
\begin{equation}
N(x,t)=\frac{1}{2|x|}H^{1,0}_{1,1}\left[\frac{|x|}{\eta^{1/2}t^{1/2}}\left|^{(1,1/2)}_{(1,1)}\right.\right]=(4\pi \eta t)^{-1/2} exp[-\frac{|x|^2}{4\eta t}],
\end{equation}
which is the classical Gaussian density.
 
In conclusion, it is seen that the solution given by (27) does not admit a probabilistic interpretation in contrast with fractional reaction-diffusion based on Caputo derivative derived by the authors (Haubold et al., 2007). However, when $\beta\rightarrow1$ , then  it has  a probabilistic interpretation, as can be seen in special cases of corollary 3.\par
\medskip
\noindent
{\bf Acknowledgment}\\
The authors would like to thank the Department of Science and Technology, Government of India, New Delhi, for the financial assistance under the project SR/S4/MS: 287/05.\par
\bigskip
\noindent
{\bf References}\par
\medskip
\noindent
Caputo, M. (1969). {\it Elasticita e Dissipazione}, Zanichelli, Bologna.\par
\smallskip
\noindent 
Feller, W. (1952). On a generalization of Marcel Riesz' potentials and the semi-groups generated by them. {\it Meddelanden Lunds Universitets Matematiska Seminarium}
(Comm. S\'{e}m.Math\'{e}m. Universit\'{e} de Lund), Tome suppl. D\'{e}di\'{e} \'{a} M.Riesz, Lund, 73-81.\par
\smallskip
\noindent 
Feller, W. (1971). {\it An Introduction to Probability Theory and its Applications}, Vol. 2, 2nd Edition, Wiley, New York (1st Edition 1966).\par
\smallskip
\noindent 
Gafiychuk, V., Datsko, B. and Meleshko, V. (2006). Mathematical modeling of pattern formation in sub- and superdiffusive reaction-diffusion systems,  arXiv: nlin. AO/0611005 v3.\par
\smallskip
\noindent 
Gafiychuk, V., Datsko, B. and Meleshko, V. (2007). Nonlinear oscillations and stability domains in fractional reaction-diffusion systems, arXiv: nlin PS/0702013v1.\par
\smallskip
\noindent
Haubold, H.J., Kumar, d., Nair, S. and Joseph, D.P. (2010). Specialfunctions and pathways for problems in astrophsysics: An essay in honor of A.M. Mathai, {\it Fractional Calculus and Applied Analysis}, {\bf 13}, 133-157.\par 
\smallskip
\noindent  
Haubold, H.J., Mathai, A.M. and Saxena, R.K. (2007). Solution of  fractional reaction-diffusion equations in terms of the H-function, Proceedings of the Second UN/ESA/NASA Workshop on the International Heliophysical Year 2007 and basic Space Science, Indian Institute of Astrophysics, Bulletin of the Astronomical Society of India, {\bf 35}, 681-689.\par
\smallskip
\noindent 
Haubold, H.J. and Mathai, A.M. (1995). A heuristic remark on the periodic variation in the number of solar neutrinos detected on Earth. {\it Astrophysics and Space Science}, {\bf 228}, 113-124.\par
\smallskip
\noindent 
Haubold, H.J. and Mathai, A.M. (2000). The fractional kinetic equation and thermonuclear functions, {\it Astrophysics and Space Science.}, {\bf 273}, 53-63.\par
\smallskip
\noindent 
Haubold, H.J. and Mathai, A.M. (2010). {\it Proceedings of the Third UN/ESA/NASA Workshop on the International Heliophysical Year 2007 and Basic Space Science}, National Astronomical Observatory of Japan, Astrophysics and Space Science Proceedings, Springer, Berlin Heidelberg.\par 
\smallskip
\noindent 
Henry, B.I. and Wearne, S.L. (2000).  Fractional reaction-diffusion, {\it Physica A}, {\bf 276}, 448-455.\par
\smallskip
\noindent 
Henry, B.I. and Wearne, S.L. (2002). Existence of Turing instabilities in a two-species fractional reaction-diffusion system, {\it SIAM Journal of Applied Mathematics}, {\bf 62}, 870-887.\par
\smallskip
\noindent 
Henry, B.I., Langlands, T.A.M., and Wearne, S.L. (2005). Turing pattern formation in fractional activator-inhibitor systems, {\it Physical Review E}, {\bf 72}, 026101.\par
\smallskip
\noindent 
Kilbas, A.A., Srivastava, H.M. and Trujillo, J.J. (2006). {\it Theory and Applications of Fractional Differential Equations}, Elsevier, Amsterdam.\par
\smallskip
\noindent 
Kilbas, A.A., Pierantozzi, T. and Trujillo, J.J. (2006a). On generalized fractional evolution-diffusion equations, published as a chapter in the book {\it Differentiation and its Applications}, A Le Hahaut\'{e},
J.A. Tenreiro, J.C. Trigeassouy, J. Sabatier (Eds.), Books on demand, Alemania (ISBN 3-8608-026-3).\par
\smallskip
\noindent 
Langlands, T.A.M., Henry, B.I. and Wearne, S.L. (2007). Turing pattern formation with fractional diffusion and fractional reactions, {\it Journal of Physics: Condensed Matter}, {\bf 19} (2007) 065115.\par
\smallskip
\noindent 
Mainardi, F. (2010). {\it Fractional Calculus and Waves in Linear Viscoelasticity: An Introduction to Mathematical Models}, Imperial College Press, London.\par
\smallskip
\noindent 
Mainardi, F., Luchko, Y. and Pagnini, G., (2001). The fundamental solution of the space-time fractional diffusion equation, {\it Fractional Calculus and Applied Analysis}, {\bf 4}, 153-192.\par
\smallskip
\noindent  
Mainardi, F., Pagnini, G. and Saxena, R.K. (2005). Fox H-functions in fractional diffusion, {\it Journal of Computational and Applied Mathematics}, {\bf 178}, 321-331.\par
\smallskip
\noindent 
Mathai, A.M. and Saxena, R.K. (1978). {\it The H-function with Applications in Statistics and Other Disciplines}, John Wiley and Sons, New York, London and Sydney.\par
\smallskip
\noindent 
Mathai, A.M. and Haubold, H.J. (2008). {\it Special Functions for Applied Scientists}, Springer, New York.\par
\smallskip
\noindent 
Mathai, A.M., Saxena, R.K. and Haubold, H.J. (2010). {\it The H-Function: Theory and Applications}, Springer, New York.\par
\smallskip
\noindent 
Miller, K.S. and Ross, B. (1993). {\it An Introduction to the Fractional Calculus and Fractional Differential Equations}, John Wiley and Sons, New York.\par
\smallskip
\noindent
Oldham, K.B. and Spanier, J. (1974). {\it The Fractional Calculus. Theory and Applications of Differentiation and Integration of Arbitrary Order}, Academic Press, New York.\par
\smallskip
\noindent
Saichev, I.A. and Zaslavasky, G.M. (1997). Fractional kinetic equations : Solutions and applications, {\it Chaos}, {\bf 7} (4), 753-764.\par
\smallskip
\noindent
Samko, S.G., Kilbas, A.A. and Marichev, O.I. (1993). {\it Fractional Integrals and Derivatives: Theory and Applications}, Gordon and Breach, New York.\par
\smallskip
\noindent
Saxena, R.K., Mathai, A.M. and Haubold, H.J.(2006a). Fractional reaction-diffusion equations, {\it Astrophysics and Space Science}, {\bf 305}, 289-296.\par
\smallskip
\noindent
Saxena, R.K., Mathai, A.M. and Haubold, H.J. (2006b). Reaction-diffusion  systems and nonlinear waves, {\it Astrophysics and Space Science}, {\bf 305}, 297-303.\par
\smallskip
\noindent
Saxena, R.K., Mathai, A.M. and Haubold, H.J. (2006c). Solution of generalized fractional reaction-diffusion equations,  {\it Astrophysics and Space Science}, {\bf 305}, 305-313.\par
\smallskip
\noindent
Uchaikin, V.V. and Zolotarev, V.M. (1999). {\it Chance and Stability: Stable Distributions and Their Applications}, VSP, Utrecht. 
\end{document}